\documentclass[twocolumn,prb,aps,notitlepage]{revtex4-1}
\usepackage{tikz}
\usepackage{amssymb}
\usepackage{CJK}
\usepackage{pdfsync}
\usepackage{amsmath}
\usepackage{graphicx,bm,epstopdf}
\usepackage{subfigure}
\usepackage{comment}
\usepackage{verbatim}
\usepackage{my_symbols}
\usepackage{color}
\usepackage[normalem]{ulem}

\begin{document}

\begin{CJK*}{UTF8}{gbsn} % Use default fonts from CJK (see below)

\title{Field-Free Synthetic-Ferromagnet Spin Torque Oscillator}
%\title{Field-free spin torque oscillator with a synthetic ferromagnetic free layer}
%+++++++++++++++++++++++++++++++++++++++++++++++++++++++++++++++++++++++++++
\author{Yan Zhou$^1$}
\author{Jiang Xiao (萧江)$^{2,3}$}
\email[Corresponding author:~]{xiaojiang@fudan.edu.cn}
\author{Gerrit E. W. Bauer$^{4,5}$}
\author{F. C. Zhang$^{1,6}$}

\affiliation{$^1$Department of Physics, The University of Hong Kong, Hong Kong, China \\
$^2$Department of Physics and State Key Laboratory of Surface Physics, Fudan University, Shanghai, China \\
$^3$Center for Spintronic Devices and Applications, Fudan University, Shanghai, China \\
$^4$Institute for Materials Research, Tohoku University, Sendai, Japan \\
$^5$Kavli Institute of NanoScience, Delft University of Technology, Delft, The Netherlands \\
$^6$Center of Theoretical and Computational Physics, Univ. of Hong Kong, Hong Kong, China
}
%#########################################################################

\date{\today}

\begin{abstract}
We study the magnetization dynamics of spin valve structures with a free composite synthetic ferromagnet (SyF) that consists of two ferromagnetic layers coupled through a normal metal spacer. A ferromagnetically coupled SyF can be excited into  dynamical precessional states by an applied current without external magnetic fields. We analytically determine the stability of these states in the space spanned by the current density and SyF interlayer exchange coupling. Numerical simulations confirm our analytical results.
\end{abstract}
%#########################################################################
\maketitle
\end{CJK*}
%#########################################################################
%\section{Introduction}
%#########################################################################
The transfer of angular momentum between the magnetic layers of current-driven spin valves (spin-transfer torque) has not been so long ago predicted  \cite{Slonczewski1996,Berger1996} and experimentally confirmed. \cite{Sun1999JMMM,Katine2000} The implied efficient electrical control of magnetizations motivated the pursue of  new research directions. When the current density exceeds a critical value, the spin-transfer torque can switch the magnetization to a different static configuration without the necessity of applied magnetic fields, which makes it attractive for next generation Magnetoresistive Random Access Memory (MRAM) application. \cite{Sun1999JMMM,Silva2008,Katine2008,Sun2008} Under an external magnetic field, the spin-transfer torque can also drive the magnetization into sustainable coherent oscillations spanning a wide frequency range from a few MHz to several hundred GHz.  \cite{Sun1999JMMM,Silva2008,Katine2008,Sun2008,Hoefer2005PRL,Kiselev2003,Tsoi1998,Tsoi2000}  High frequency magnetic oscillations  generate a coherent microwave voltage signal through the Giant Magnetoresistance (GMR) in metallic spin valves or through the Tunneling Magnetoresistance (TMR) in magnetic tunnel junctions (MTJs). This effect can be used in so-called spin-torque oscillators (STO), which has many advantages including wide tunability,  \cite{Bonetti2009} very high modulation rates, \cite{Pufall2005,Muduli2011a} compact device size, and high compatibility with standard CMOS processes. \cite{Akerman2005,Engel2005} Thus STO is appealing for high frequency microwave applications including microwave emitters, modulators and detectors. \cite{Muduli2011} However, the necessity of an applied magnetic field  up to ${\sim}$1 Tesla has greatly limited the potential of these STOs for microwave generation and wireless communication applications. Recently, various solutions have been proposed to enable zero-field operation, \textit{viz}. STO with a perpendicularly magnetized fixed \cite{Houssameddine2007} or spin valves with out-of-plane magnetized free layer, \cite{Rippard2010,Mohseni2011} magnetic vortex oscillators, \cite{Pribiag2007,Pribiag2009,Finocchio2010,Locatelli2011,Dussaux2010,Dussaux2011} wavy-torque STO by judicially choosing free and fixed layer materials with different spin diffusion lengths, \cite{Boulle2007} and  a tilted magnetization of the fixed layer with respect to the film plane. \cite{Zhou2009c,Zhou2008,Zhou2009d,Wang2011,He2010}

 Recently, synthetic ferromagnets (SyFs) composed of two ferromagnetic layers separated by a very thin nonmagnetic spacer have been used to replace the  free layer of a spin valve or MTJ. \cite{Taniguchi2011,Yakata2009,Yulaev2011,Ichimura2011,Balaz2011,Bergman2011,Klein2012} SyF based spintronic devices have the advantage of higher thermal stability, smaller stray magnetic fields, faster switching speed and reduced threshold switching current as compared to single ferromagnetic free layers. \cite{Taniguchi2011,Yakata2009,Yulaev2011,Ichimura2011,Balaz2011,Bergman2011,Klein2012}  Klein \etal \cite{Klein2012} predicted that an anti-ferromagnetically coupled SyF layer with uncompensated magnetization can generate microwave oscillations at zero applied magnetic field. 

 Here we predict that a \textit{ferromagnetically} coupled SyF can also be driven into dynamical precessional states, which, however, are surrounded in parameter space by static canted states with non-collinear magnetizations. We use an analytical approach to determine the stability regimes of the SyF system  and confirm results by numerical simulations.
%Using our model, we can perform fast calculations (a few seconds) to determine the parameter space in which strong spin torque oscillator states of SyF systems exist. Furthermore, we propose an effective synchronization mechanism that can phase-lock large network of STO to achieve enhanced power output. The synchronization scheme can be easily studied an analytical model facilitated by numerical simulations.

%The manuscript is organized as follows. Section \ref{Analytical model} describes the analytical model we used to study the composite SyF system. Section \ref{Spin-torque driven magnetization procession at zero applied field} presents the comparison between analytical results and macrospin simulations. Section \ref{Discussions and conclusions}  discusses  recent studies in the literature relevant to the current work and summarizes the main results.

%\section{Analytical model}
%\label{Analytical model}

%-----------------------------
\begin{figure}[b]
\centering
      \includegraphics[width=0.46\textwidth]{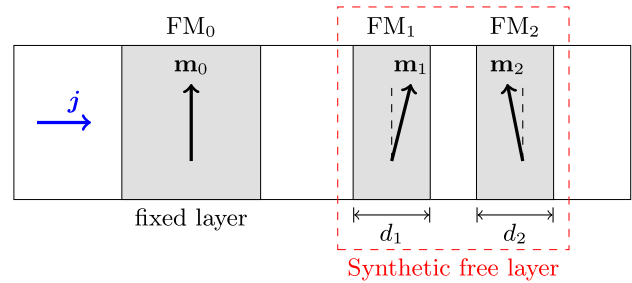}
%\begin{tikzpicture}[scale=1]
%        \draw[very thick, ->] (0.3,1) -- (1,1) node[above left,blue] {${\bm j}$} [color=blue];
%        \draw[very thin] (0,0) rectangle (8,2);
%
%        \draw[very thin] (1.4,0) rectangle (3.2,2) [fill=gray!30!white];
%        \draw[very thick, ->] (2.3,0.5) -- (2.3,1.5) node[above] {$\mm_0$};
%        \node[above] at (2.3,2) {FM$_0$};
%        \node[below] at (2.3,0) {fixed layer};
%
%        \draw[very thin] (4.4,0) rectangle (5.4,2) [fill=gray!30!white];
%        \draw[dashed,-] (4.9,0.5) -- (4.9,1.5);
%        \draw[very thick, ->] (4.9,0.5) -- (5.15,1.5) node[above] {$\mm_1$};
%        \node[above] at (4.9,2) {FM$_1$};
%        \draw[|<->|] (4.4,-0.2) -- (4.9,-0.2) node[below] {$d_1$} -- (5.4,-0.2);
%
%        \draw[very thin] (6.0,0) rectangle (7.0,2) [fill=gray!30!white];
%        \draw[dashed,-] (6.6,0.5) -- (6.6,1.5);
%        \draw[very thick, ->] (6.6,0.5) -- (6.4,1.5) node[above] {$\mm_2$};
%        \node[above] at (6.5,2) {FM$_2$};
%        \draw[|<->|] (6.0,-0.2) -- (6.5,-0.2) node[below] {$d_2$} -- (7.0,-0.2);
%
%        \draw[very thin, dashed] (4.2,-0.65) rectangle (7.2,2.5) [color=red];
%        \node[red, below] at (5.7,-0.65) {Synthetic free layer};
%
%\end{tikzpicture}
    \caption{A spin valve structure with an SyF free layer, where FM$_0$ is the fixed layer and FM$_{1,2}$ layers are (anti-) ferromagnetically coupled.}
\label{fig:spin_valve_SyF}
\end{figure}
%-----------------------------
%%-----------------------------
%\begin{figure}[b]
%%      \includegraphics[width=0.45\textwidth,trim= 0 0 0 0, clip=true]{syf_sch}
%%\includegraphics[scale=0.45, clip=true, viewport=0in 0in 10in 3in]{syf_sch.eps}
%\includegraphics[width=0.45\textwidth,trim= 0 0 0 0, clip=true]{syf_sch}
%       \caption{A spin valve structure with an SyF free layer, where FM$_0$ is the fixed layer and FM$_{1,2}$ layers are (anti-)ferromagnetic coupled and form an SyF free layer.}
%       \label{fig:spin_valve_SyF}
%\end{figure}
%%-----------------------------

We study a spin torque nanodevice with synthetic ferromagnetic free layers as shown in \Figure{fig:spin_valve_SyF}. The left ferromagnetic film forms the fixed polarizer with magnetization $\mm_0\|\hzz$, and the SyF consists of two ferromagnetic layers FM$_1$ and FM$_2$ of thickness $d_{1,2}$ with a paramagnetic spacer. The unit vectors describing the magnetization orientation are $\mm_1$ for FM$_1$ and $\mm_2$ for FM$_2$. For simplicity, we assume that the SyF layers are made of the same materials with identical saturation magnetization $M_s$. The exchange coupling strength reads $E_C = - JS \ml{\cdot}\mr$, where $J$ and $S$ are the coupling energy per unit area and the cross section area of the sample, respectively. This corresponds to an effective coupling field $\HH_i^{\rm c} = Jm_{\bar{i}}/({\mu}_0M_sd_{i})$, where $i=1,2$ and $\bar{i}=3-i$, ${\mu}_0$ is the vacuum magnetic susceptibility. $\mm_1$ and $\mm_2$ can be parallel or anti-parallel at zero applied field, corresponding to the non-local Ruderman-Kittel-Kasuya-Yoshida (RKKY) exchange ferromagnetic ($J>0$) or antiferromagnetic ($J<0$) coupling, respectively. The spacer between FM$_0$ and FM$_1$ is presumed thick enough that the RKKY coupling with the fixed layer is negligibly small.
Although the dynamic dipolar coupling may be responsible for the apparent reduction of static magnetization\cite{Dmytriiev2010} or linewidth of the current-induced spin wave mode\cite{Gusakova2011}, it is estimated to be much smaller for our case compared to the shape anisotropy field and the other fields due to current-induced spin torque and interlayer exchange coupling and therefore disregarded \cite{Klein2012}.

Let $P_{0,1}$ be the spin current polarization by $\mm_{0,1}$ such that the spin current density in the two spacers are $P_0j$ and $P_1j$ with $j$ the electric current density. The corresponding spin-transfer torques on $\ml$ and $\mr$ are given by the projections:
%-----------------------------------------
\begin{subequations}
\label{eq_stt}
\begin{align}
        \NN_{\rm ST1} &= \frac{{\gamma}{\hbar} j}{2 e {\mu}_{0}M_{s}d_1}\ml{\times}(P_0\mm_0-P_1\mr){\times}\ml, \\
        \NN_{\rm ST2} &= \frac{{\gamma}{\hbar} j}{2 e {\mu}_{0}M_{s}d_2}P_1\mr{\times}\ml{\times}\mr,
\end{align}
\end{subequations}
%-----------------------------------------
with ${\gamma}$ the gyromagnetic ratio and $P_0$ ($P_1$) are in general functions of the angle ${\theta}={\angle}(\mm_0,\mm_1)$ $({\angle}(\mm_1,\mm_2))$. \cite{Slonczewski1996, Xiao2004}

Spin pumping  causes enhanced damping in a ferromagnetic layer by emitting spin current into the adjacent non-magnetic layers. \cite{Tserkovnyak2002} This emitted spin pumping current can exert a torque on the second layer. Disregarding the backflow and diffusion in the spacer layer, the torque density acting on $\mm_i$ due to spin pumping from $\mm_{\bar{i}}$ can be written as
%-----------------------------
\begin{align}
\label{eq_spin pumping}
        \NN_{\rm SPi} ={\beta}\mm_{\bar{i}} {\times} \dmm_{\bar{i}}
        -[({\beta}\mm_{\bar{i}} {\times} \dmm_{\bar{i}}){\cdot}\mm_{{i}}]\mm_{{i}}
\end{align}
%-----------------------------
where ${\beta}$ is the effective enhanced damping due to spin pumping. It has been shown that \Eq{eq_spin pumping} gives rise to a dynamic exchange interaction that can induce synchronization of the magnetization dynamics in two neighboring ferromagnetic layers even for wide spacers. \cite{Tserkovnyak2005} In the results below we fully include the spin pumping. However, in contrast to multilayers excited by microwaves, \cite{Tserkovnyak2005} we observe here only small corrections demonstrating the dominance of charge current-induced torques.

The dynamics is described by the coupled Landau-Lifshitz-Gilbert-Slonczewski (LLGS) equations, \cite{sunjz2000,Xiao2005}
%-----------------------------
\begin{equation}
\label{eq_LLGS}
\dmm_i = - {\gamma} \mm_i {\times} \HH_i+ {\alpha} \mm_i {\times} \dmm_i -\NN_{\rm SPi}-\NN_{\rm STi},
\end{equation}
%-----------------------------
where ${\alpha}$ is the sum of the intrinsic Gilbert  and the spin pumping induced damping. \cite{Tserkovnyak2005} The effective magnetic fields $\HH_i$ consist of shape anisotropy and RKKY exchange coupling and can be written as,
%-----------------------------
\begin{equation}
\label{eq_Heff}
\HH_{i} = \frac{2K_u}{{\mu}_0M_s}[\mm_i{\cdot}\ee_z]\ee_z+\frac{J\mm_{\bar{i}}}{{\mu}_0M_sd_{i}}.
\end{equation}
%-----------------------------

For simplicity, we consider $d_1=d_2=d$ (equal magnetization) for the rest of the paper unless otherwise specified. We linearize  \Eq{eq_LLGS} in the vicinity of four collinear equilibrium states, \ie $\up\up, \up\dn, \dn\up, \dn\dn$, and assume $\mi = {\lambda}_i\hzz+\uu_i $ with  ${\lambda}_i={\pm}$ and $\uu_i$ denoting the small transverse magnetization component. After the linearization and the Fourier transform $\uu_i(t) = {\int} \td{\uu}_i({\omega})e^{-i{\omega} t}d{\omega}/2{\pi}$, \Eq{eq_LLGS} becomes
%%-----------------------------
%\begin{subequations}
%\label{eq_lin3}
%\begin{align}
%(1-i{\alpha}{\lambda}_1){\omega}\tilde{\uu}_1&=-({\lambda}_1 {\omega}_0+{\lambda}_2 {\omega}_J)\tilde{\uu}_1+{\lambda}_1{\omega}_J \tilde{\uu}_2 \\
%&%+ i{\alpha}{\lambda}_1{\omega}\tilde{\uu}_1
%-i{\beta}{\lambda}_2{\omega}\tilde{\uu}_2+i{\omega}_j[\tilde{\uu}_2+{\lambda}_1(1-{\lambda}_2)\tilde{\uu}_1], \nn
%(1-i{\alpha}{\lambda}_2){\omega}\tilde{\uu}_2&=-({\lambda}_2 {\omega}_0+{\lambda}_1 {\omega}_J)\tilde{\uu}_2+{\lambda}_2{\omega}_J \tilde{\uu}_1 \\
%&%+ i{\alpha}{\lambda}_2{\omega}\tilde{\uu}_2
%- i{\beta}{\lambda}_1{\omega}\tilde{\uu}_1-i{\omega}_j[\tilde{\uu}_1-{\lambda}_1{\lambda}_2\tilde{\uu}_2] \nonumber
%\end{align}
%\end{subequations}
%%-----------------------------
%-----------------------------
\begin{equation}
\label{eq_AV}
        \smlb{\hat{A}{\omega}+\hat{V}}
        \smlb{\begin{array}{c} \tilde{\uu}_1 \\ \tilde{\uu}_2\end{array}} = 0
\end{equation}
%-----------------------------
with
%-----------------------------
\begin{subequations}
\begin{align}
        \hat{A} &=
        \smatrix{1-i{\alpha}{\lambda}_1 & i{\beta}{\lambda}_2 \\ i{\beta}{\lambda}_1 & 1-i{\alpha}{\lambda}_2}, \\
        \hat{V} &=
        {\omega}_0\smatrix{ {\lambda}_1 & 0 \\ 0 & {\lambda}_2 }
        + {\omega}_J\smatrix{ {\lambda}_2 & -{\lambda}_1 \\ -{\lambda}_2 & {\lambda}_1 } \nn
        &+ i{\omega}_j\midb{P_0\smatrix{-{\lambda}_1 & 0 \\ 0 & 0 } 
        + P_1\smatrix{{\lambda}_1{\lambda}_2 & - 1 \\ 1 & -{\lambda}_1{\lambda}_2 } },
%        \smlb{\begin{array}{cccc}
%                {\lambda}_1{\lambda}_2{\omega}_j-{\lambda}_1{\omega}_j-i({\lambda}_1{\omega}_0+{\lambda}_2{\omega}_J) & - {\omega}_j + i{\lambda}_1{\omega}_J \\
%               {\omega}_j+i{\lambda}_2{\omega}_J & -{\lambda}_1{\lambda}_2{\omega}_j-i({\lambda}_2{\omega}_0+{\lambda}_1{\omega}_J)
%        \end{array}}
\end{align}
\end{subequations}
%-----------------------------
with ${\omega}_0=2{\gamma}K_u/{\mu}_0M_s$, ${\omega}_J={\gamma}J/{\mu}_0M_sd$, ${\omega}_j=({\hbar}/2e)({\gamma}j/{\mu}_0M_s d)$.
The frequency of the normal modes are given by the eigenvalues of $\hat{W} = -\hat{A}^{-1}\hat{V}: {\Omega}_1$ and ${\Omega}_2$.
%%-----------------------------
%\begin{equation}
%\label{eq_w}
%        \hat{W} = -\hat{A}^{-1}\hat{V}: \quad {\Omega}_1 \qand {\Omega}_2.
%\end{equation}
%%-----------------------------
When any of the $\im{{\Omega}_{1,2}} > 0$, the system is unstable, implying that an infinitesimal perturbation will lead to magnetization dynamics with amplitudes that initially increase exponentially in time.

%#########################################################################
%\section{Spin-torque driven magnetization procession at zero applied field}
%\label{Spin-torque driven magnetization procession at zero applied field}
%\subsection{Symmetric case $d_{1}=d_{2}$}

%\subsection{Intermediate coupling strength}

\begin{figure}[t]
        \includegraphics[width=0.48\textwidth,trim= 0 0 0 0, clip=true]{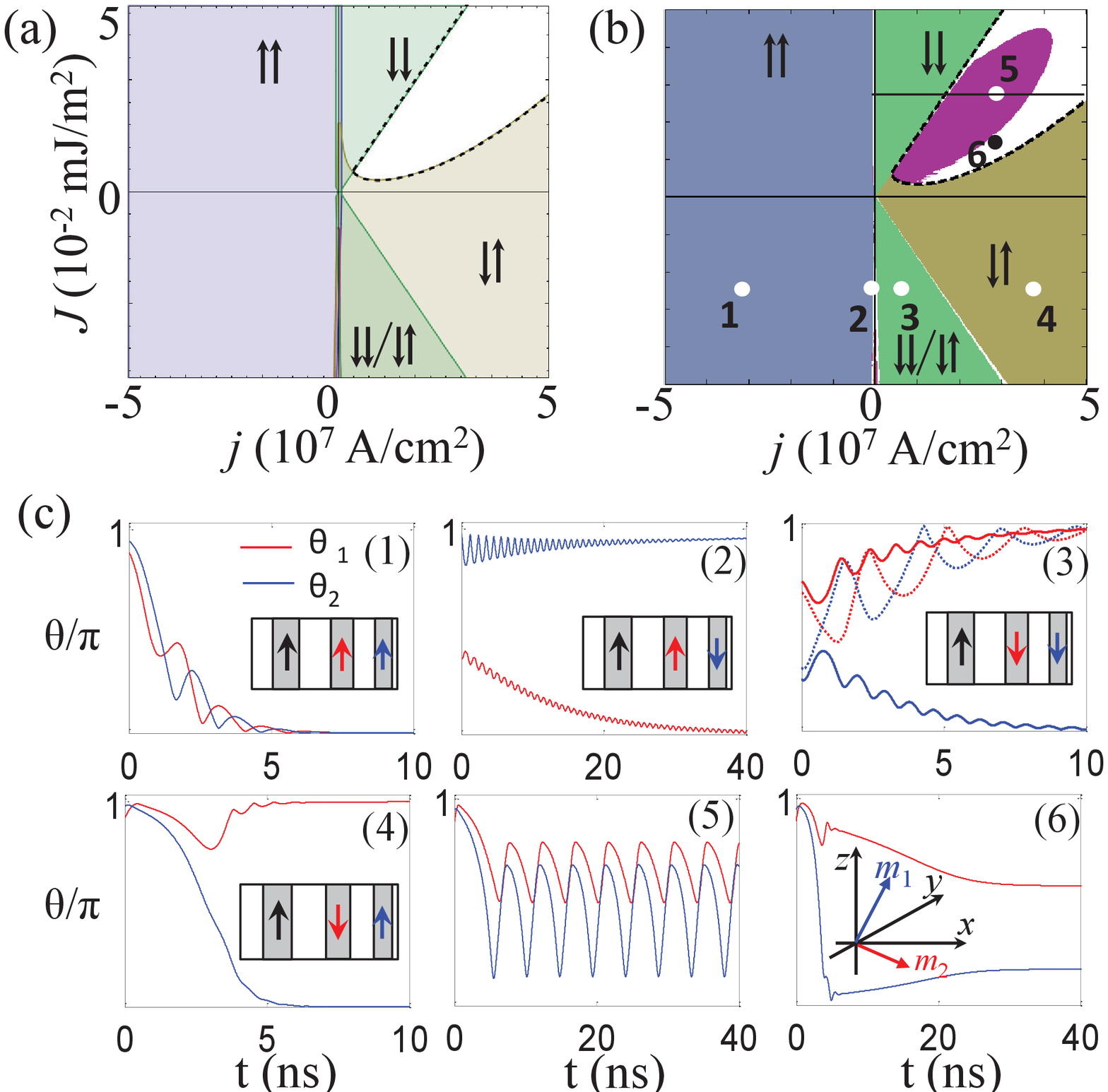}
        \caption{(Color online) Dynamical phase diagram in the parameter space of currents and RKKY coupling strengths. (a) Phase diagram calculated analytically by \Eq{eq_AV}; none of the four states $\up\up, \up\dn, \dn\up, \dn\dn$ is stable in the white region. (b) Phase diagram calculated by numerically solving the LLGS \Eq{eq_LLGS}. The purple are  the STO phase, and the white one the canted state. (c) The time evolution of the polar angles ${\theta}_{1,2} = {\angle}(\mm_{1,2},\hzz)$ at the six different points indicated in the phase diagram. In the third subfigure of (c), the solid and dashed lines correspond to different sets of initial conditions.}
%Points 1-4 correspond to $j=-2.4{\times}10^7$  A/cm$^2$, $-0.04{\times}10^7$  A/cm$^2$, $0.76{\times}10^7$  A/cm$^2$, 3.32${\times}10^7$  A/cm$^2$, and the same $J=-3{\times}$10$^{-2}$ mJ/m$^2$. Point 5 corresponds to  $j=2.4{\times}10^7$  A/cm$^2$ and $J=3{\times}$10$^{-2}$ mJ/m$^2$. Point 6 corresponds to $j=2.4{\times}10^7$  A/cm$^2$ and $J=1{\times}$10$^{-2}$ mJ/m$^2$.
        \label{fig:phase_diagram}
\end{figure}

The above results allow us to calculate the stability regions for the $\up\up, \up\dn, \dn\up, \dn\dn$ phases in the  space of typical experimental parameters:  angle-independent  $P_1 = P_2 = P = 0.5$, $d =3$~nm, $K_u =8{\times}10^{4}$ J/m$^3$, $j{\sim}10^8$ A/cm$^2$ and $J{\sim}1$ mJ/m$^2$. \cite{Klein2012,Yakata2010} To analytically construct the stability diagram as shown in the top-left panel of \Figure{fig:phase_diagram}(a), we first calculate the eigenvalues for each given set of [$j, J$] as given by \Eq{eq_AV}. Then we determine whether any of the four collinear static states (different combinations of [${\lambda}_1,{\lambda}_2$]) is stable or not. For example,  both the imaginary part of the eigenvalues of $\up\up$ configuration [${\lambda}_1=+1,{\lambda}_2=+1$] are negative when $j{\lesssim}0$. Therefore $\up\up$  is stable in the blue region. In this way, we quickly map the parameter space for any given set of [$j, J$] and construct the entire stability diagram consisting of four collinear magnetization configurations. The spin torque drives the SyF to the parallel  $\up\up$ configuration for negative currents $j$. For positive currents, the $\dn\up$ configuration is preferred. These results can be understood from \Eq{eq_stt}. In a small region  the antiparallel $\up\dn$ state exists for negative $J$ and small $j$ (\ie in the vicinity of the negative vertical axis but not visible in the figure due to the scale). Although it seems that the $\dn\dn$ state also occupies the fourth quadrant ($j>0, J<0$), this triangular region is hysteretic, \ie $\dn\dn$ and $\dn\up$ may both appear depending on the history.

%-----------------------------
%-----------------------------

Most importantly, there is a white/purple region in which none of the four static collinear states is stable, therefore  it must be either in  a dynamical STO or static canted state. To leading order of ${\alpha}$ , we find from \Eq{eq_AV} an approximate boundary for the white region:
%-----------------------------
\begin{subequations}
\label{eq_boundary}
\begin{align}
        \mbox{upper: }& {\omega}_J = {\omega}_j,\\
        \mbox{lower: }& {\omega}_J = \sqrt{4{\omega}_0^2+{\omega}_j^2} - 2{\omega}_0 + {\alpha}{{\omega}_0^2\ov{\omega}_j},
%    &wj= \begin{cases}
%        {2{\alpha}({\omega}_0-{\omega}_J)\sqrt{{\omega}_0({\omega}_0-2{\omega}_J)}\ov {\omega}_0+{\omega}_J-\sqrt{{\omega}_0({\omega}_0-2{\omega}_J)}} & {\omega}_j {\ll} {\omega}_0, \\
%        \sqrt{{\omega}_J(4{\omega}_0+{\omega}_J)} &\mbox{otherwise.}
%    \end{cases}
%       &{\omega}_j = 2{\alpha}({\omega}_0-{\omega}_J) +\im{\sqrt{(2{\omega}_0-i{\omega}_j)(2{\omega}_0-4{\omega}_J+3i{\omega}_j)}} \\
\end{align}
\end{subequations}
%-----------------------------
which is plotted as the black dashed lines in \Figure{fig:phase_diagram}(a,b), matching the numerically obtained boundaries almost exactly. \Eq{eq_boundary} is calculated from the eigenvalue analysis based on \Eq{eq_AV} with perturbation from the four static collinear states. This method is equivalent to that used by Bazaliy \textit{et al.} \cite{Bazaliy2004PRB}  A fully analytical solution for the boundary between STO and static canted  phase turned out to be intractable. due to the complexity of \Eq{eq_AV} for non-collinear states.  

% We have to point out that although we indicated the non-white region in the phase diagrams in \Figure{fig:phase_diagram} as one or more of the four collinear states ($\up\up, \up\dn, \dn\up, \dn\dn$), we do not exclude the possibility that there might be some STO or canted states within the non-white region when all possibile initial states are tried.

We now present numerical solutions of the LLGS \Eq{eq_LLGS} including  damping, spin torque and RKKY coupling. We summarize the  dynamics of the coupled $\mm_1$ and $\mm_2$ in \Figure{fig:phase_diagram}(b), in which we confirm the phase boundaries in the analytical analysis in \Figure{fig:phase_diagram}(a). In addition, we can now map the STO phase by the purple color. The rest of the white region consists of static canted states. In \Figure{fig:phase_diagram}(c) we show the six different SyF configurations that may exist depending on the current and RKKY coupling strength. Point 5 corresponds to an STO state, in which both $\mm_{1,2}$ are undergoing large angle precessions, which result in a large magnetoresistance oscillations attractive for applications.

\begin{figure}[t]
\includegraphics[width=0.48\textwidth,trim= 0 0 0 0, clip=true]{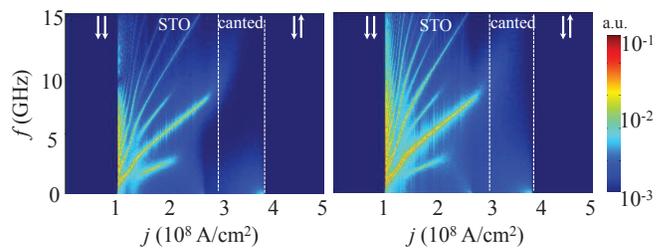}
        \caption{(Color online) Power spectrum for $\mm_0{\cdot}\mm_1$ (left) and $\mm_1{\cdot}\mm_2$ (right) as a function of current density $j$ and frequency $f$ at $J=0.25$ mJ/m$^2$,  corresponding to the black line in the top right panel of \Figure{fig:phase_diagram}.} 
\label{fig:power_spectrum}
\end{figure}

For the STO phase, we study the power spectrum of the magnetoresistance due to the magnetization oscillation of $\mm_{1,2}$, which is  approximated by $R(t) = R_0 + {\Delta}R_1\mm_0{\cdot}\mm_1 + {\Delta}R_2\mm_1{\cdot}\mm_2$. \Figure{fig:power_spectrum} shows the Fourier transform of $\mm_0{\cdot}\mm_1$ (left) and $\mm_1{\cdot}\mm_2$ (right) as a function of current density $j$ at $J=0.25$ mJ/m$^2$ (corresponding to the black line in \Figure{fig:phase_diagram}(b)). The clear  higher order harmonic modes are evidence of the non-linearities in the STO dynamics. \Figure{fig:power_spectrum} also demonstrate that the oscillation frequency of the device can be continuously tuned by the current at zero applied magnetic field and thus potentially be utilized for  nano-scale microwave applications.
It should be noted that the frequency range can be further tuned by tens of GHz by adopting a larger $K_u$ or taking into account the easy-plane anisotropy field (demagnetization field).
%-----------------------------
%-----------------------------

%\section{Discussions and conclusions}
%\label{Discussions and conclusions}

The STO phase studied in this work differs from that studied by Klein \etal  [\onlinecite{Klein2012}]. The STO phase found by Klein \etal arises only in an {\it anti-ferromagnetically} ($J < 0$) coupled {\it uncompensated} SyF ($M_1=M_sd_1S<M_sd_2S=M_2$), in which the total magnetization for the SyF is opposite to that of $\mm_0$. However, the STO phase found in our study appears in the ferromagnetically coupled SyF with $J > 0$ and does not require $M_1{\neq} M_2$. Furthermore, we were not able to reproduced the STO phase found  by Klein \textit{et al}. for an uncompensated and antiparallel SyF. 
% This discrepancy may be due to  the different treatment of spin transport. We assume angle-independent $P_i$'s, leading to torques that are governed by geometrical projections of the magnetization angles.  In Ref. [\onlinecite{Klein2012}] spin diffusion and the spin accumulations are taken into account, which affect the angle dependence of the current-induced torques.
We  checked the effect of an angular dependence of the prefactor $P_i$  that take into account the effects of a spin accumulation \cite{Xiao2004}. The boundaries of the white region will shift noticeably, but we find no qualitative changes. The differences with Ref. \onlinecite{Klein2012} might be due to other details in handling spin transport.%Nevertheless, we both find that large-amplitude coherent spin torque driven microwave oscillations can be achieved without applied fields for a wide parameter range.

Finally, we note that our approach can be readily extended from bi-layer to multilayer systems in which each layer is exchange-coupled with its neighbouring layers (unpublished). This may provide a novel route to effectively synchronize a large network of spin torque oscillators. 

In conclusion,
%we studied spin valves with synthetic ferromagnetic free layers composed of two exchange-coupled ferromagnetic layers separated by a non-magnetic spacer layer.
we predict that the ferromagnetically coupled SyF can be driven into STO states without the need of applying magnetic fields. The resulting STO states display large angle precession, therefore generating a large power output.  In addition to dynamical STO states, static canted states are also possible in the same structure at slightly different applied current densities. Our findings may guide the experimental effort towards the field-free STO for real applications.

%We use an analytical model based on nonlinear system theory to map out the phase diagram of the system for a wide range of current density and RKKY coupling. Based on this model, we can determine the stability regions for different SyF configurations and infer the existence of current-induced oscillations. The analytical model has great advantage over micromagnetic simulations since it relaxes the need for complete magnetodynamic simulations. We have also carried out numerical simulations for the same system confirming the validity of the analytical model. The analytical eigenvalue analysis is useful for device design and optimization of spin torque nano devices. We hope our study will stimulate further experimental efforts toward this direction and can serve as a guideline for optimizing such devices.

We acknowledge support from University Research Committee (Project No. 106053) of HKU, the University Grant Council (AoE/P-04/08) of the government of HKSAR, the National Natural Science Foundation of China (No. 11004036, No. 91121002), the FOM foundation, DFG
Priority Program SpinCat, and EG-STREP MACALO.

%#########################################################################
\bibliographystyle{apsrev}
\bibliography{SyF2d}

\end{document}